\newfont {\xx} {cmti10}

\newcommand{\bea}{\begin{eqnarray}}
\newcommand{\eea}{\end{eqnarray}}
\def\slash{\hspace*{-.14in}/ \hspace*{.08in}}
\def\ss{\hspace*{-.08in}/}

\documentstyle[epsf,epsfig,12pt]{article}
\catcode`\@=11  
\@addtoreset{equation}{section}
\def\theequation{\arabic{section}.\arabic{equation}}

\begin{document}
\pagestyle{empty}
\begin{flushright}
{CERN-TH/97-192}\\
{SHEP-97-16}\\
\end{flushright}
\vspace*{5mm}
\begin{center}
{\large {\bf Calculations of One-Loop Supersymmetric Corrections to
Large-$E_T$ Jet Cross Sections}}\\
\vspace*{1cm}
{\bf John Ellis}\\
\vspace{0.3cm}
Theoretical Physics Division, CERN \\
1211 Geneva 23, Switzerland\\
and \\
{\bf Douglas A. Ross} \\
\vspace{0.3cm}
Physics Department,\\
University of Southampton,\\
Southampton SO17 1BJ, United Kingdom\\
\vspace*{2cm}
{\bf ABSTRACT} \\ \end{center}
\vspace*{5mm}
\noindent
We have calculated the one-loop supersymmetric corrections to
the $2 \to 2$ parton scattering subprocesses ${\bar q} q \to {\bar q} q,
{\bar q} q \to g g, g q \to g q$ and $g g \to g g$, including form-factor
corrections and box diagrams with internal squarks and gluinos of
arbitrary mass. In general, these exhibit cusps at the corresponding
direct-channel sparticle thresholds. We use these calculations to
make numerical estimates of the possible threshold effects at the
Fermilab Tevatron collider and at the LHC, which depend on the
rapidity range selected, but can be as large as a few percent.
These effects are diluted in the integrated large-$E_T$
cross section, where they are negative.
 
\vspace*{3cm}

\begin{flushleft}
CERN-TH/97-192\\
SHEP-97-16 \\
August  1997
\end{flushleft}
\vfill\eject

\setcounter{page}{1}
\pagestyle{plain}

\section{Introduction}

Large samples of inclusive hadronic jets are available now from
the Fermilab Tevatron collider, and will become available from
the LHC in the future. These offer the prospects of precision
tests of QCD, which is of interest and importance in its own right. 
Moreover, as the dynamics of QCD is better understood, it becomes a sharper
tool for probing possible physics beyond the Standard Model.
Complete calculations of jet physics to next-to-leading order 
in perturbation theory exist in pure QCD~\cite{KEetal,KS}. However, this 
may not be sufficient for precision physics at an
exploratory  machine that opens up a new energy range where there
may be thresholds for new physics. For instance, the Fermilab Tevatron
collider has already crossed the threshold for ${\bar t} t$
production, and either it or the LHC should cross the threshold for
squark and gluino production. In the neighbourhood of, and beyond,
the threshold for such new heavy strongly-interacting particles,
their virtual effects should be included in a complete treatment
of one-loop perturbative QCD effects.

Our attention was drawn to this problem by recent measurements of
large-$E_T$ jet cross sections at the Tevatron collider~\cite{CDF,D0}, 
some of which exhibited a {\it prima facie} discrepancy with predictions
based on the parton distribution functions available previously.
Clearly there are uncertainties in these distributions~\cite{huston,GMRS}, 
which 
are not (yet) calculable from first principles, and it has been argued in
particular~\cite{huston} that reasonable uncertainties in the gluon 
distribution
could accommodate simultaneously both the CDF~\cite{CDF} and D0~\cite{D0} 
jet measurements.
However, it has also been argued that strong-interaction threshold
effects could be significant~\cite{Betal,KW}, with emphasis placed on the
possible importance of the sparticle threshold~\cite{KW}. 
One might consider whether such a threshold effect could provide
an alternative signature for sparticles, if their decays differ
from those anticipated in conventional direct searches. Even if one 
does not expect the sparticle threshold to be very important, it is
clearly desirable to have an exact one-loop treatment
of it, just as one-loop sparticle corrections are known and
used in the analysis of precision electroweak physics at and
around the $Z^0$ peak~\cite{Hollik}.

We have undertaken detailed calculations of such corrections 
in the Minimal Supersymmetric extension of the Standard Model (MSSM), and
published first results~\cite{thin}. We found that there
could be threshold structures of the order of a few \% in the
cross sections for individual parton-parton scattering
subprocesses when one-loop corrections were included. However,
these were of the wrong shape and of insufficient 
magnitude to make a significant contribution to the
resolution of the CDF large-$E_T$ conundrum~\cite{CDF}.
The purpose of this paper is to complete the analysis
of~\cite{thin}, presenting more details of the calculations,
presenting the one-loop corrections to all partonic subprocesses,
including box diagrams and the $gg \rightarrow gg$
subprocess, which were not included in~\cite{thin}, and
combining our calculations in a complete one-loop
numerical analysis~\footnote{Numerical results of such a
study have also been published in~\cite{Alam}.} of jet cross 
sections at 
Fermilab and the LHC, including the appropriate convolutions over all
the contributing parton-parton scattering subprocesses.

The layout of this paper is as follows. 
In section 2
we present details of our results~\cite{thin} for the
calculations of the one-loop corrections to 
the partonic subprocesses arising from the form-factor corrections
to the triple-gluon vertex and the quark-quark-gluon vertex,
which include self-energy corrections due to squarks~\footnote{We assume 
for simplicity that all the relevant left- and right-handed squarks 
have equal masses. This means that our results should not be
assumed to apply to processes involving both sbottom and stop squarks.} and 
gluinos. We comment on various consistency checks on our calculations,
demonstrating in particular cancellations between different
diagrams associated with Ward identities, and remaining 
logarithmic singularities that reflect the expected running of the
strong coupling $\alpha_s$ above the supersymmetric
threshold. Section 3 contains a similar analysis of
one-loop box diagrams, including a number of further
consistency checks. Numerical results are presented
in section 4, including the convolutions with
parton distribution functions which are appropriate
for the analysis of jet cross sections at the Fermilab
Tevatron collider when $E_T \simeq m \equiv m_s = 
m_g = 200$ GeV (where $m_{s,g}$ denote the squark and gluino masses), and at 
the LHC when $E_T \simeq m = 1$ TeV. Section 5 summarizes our
conclusions from our calculations.  In the appendix we list the 
Veltman-Passarino~\cite{VP}
functions, in terms of which our results are displayed.

\section{Form-Factor Corrections}
In this section present the form-factor corrections
for the partonic subprocesses
\begin{equation} q(p_1)+\bar{q}(p_2) \to q(p_3)+\bar{q}(p_4)
\label{qqqq} \end{equation}
\begin{equation} q(p_1)+\bar{q}(p_2) \to g(p_3)+g(p_4)
\label{qqgg} \end{equation}
\begin{equation} g(p_1)+g(p_2) \to g(p_3)+g(p_4).
\label{gggg} \end{equation}
calculated using the $\overline{MS}$ prescription. The form factors for all 
other partonic 
subprocesses can be obtained from these by exploiting crossing symmetry.

\begin{figure}
\begin{center}
\leavevmode
\hbox{\epsfxsize=4.0 in
\epsfbox{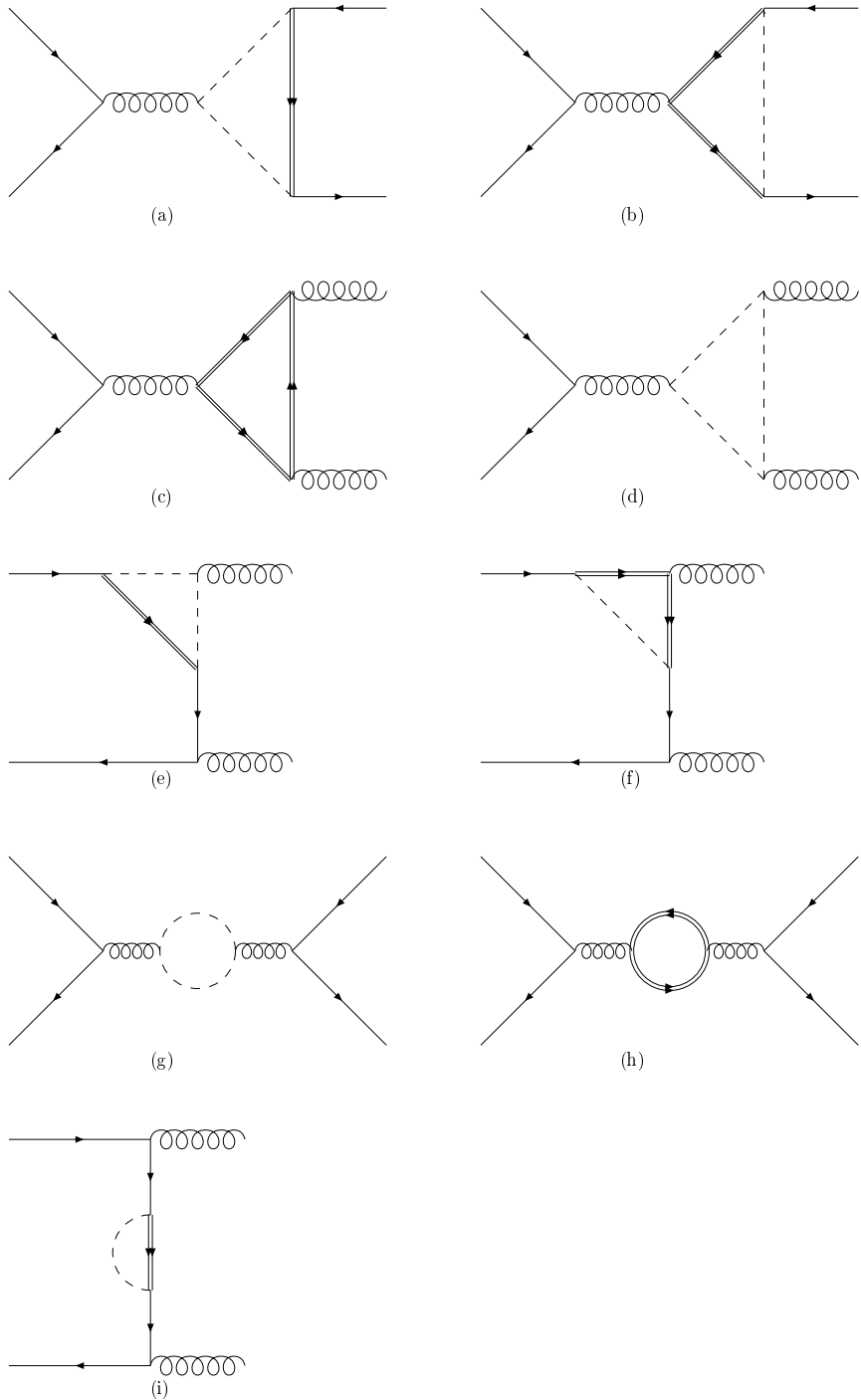}}
\end{center}
 \vspace*{-.5in}
\caption[]{One-loop Feynman diagrams involving virtual
sparticles in the MSSM for
(a,b) the virtual-$g \bar q q$ vertex,
(c,d) the virtual-$g g g$ vertex,
(e,f) the virtual-$q \bar q g$ vertex,
(g,h) the gluon self energy, and
(i) the quark self energy.
Here and in subsequent figures of Feynman diagrams,
 the broken lines represent squarks
and the double solid lines represent gluinos.} \label{axofig}
\end{figure}
 
The contribution to the amplitude for the subprocess (\ref{qqqq})
from all diagrams except box graphs may be written in the general form
\begin{equation}
\frac{4\pi i\alpha_s}{s}
 \bar{v}(p_4)F_{qqg}(s)\gamma^\mu \tau^a u(p_3)
 \,  \bar{v}(p_2)F^{qqg}(s)\gamma^\mu \tau^a u(p_1),
\label{aqqqq}
\end{equation}
where $\tau^a$ is the generator of colour $SU(3)$ in the
defining representation, and $F_{qqg}(s)$ is a form factor for the
quark-quark-virtual-gluon vertex. The tree-level amplitude is obtained by 
replacing this form factor by unity.
The one-loop corrections to this form factor from internal squarks and 
gluinos
come from the diagrams shown in Fig.~\ref{axofig}(a,b,g,h), and give a contribution to
$F_{qqg}$ of the form
\begin{eqnarray}
\Delta F_{qqg}&=& \frac{\alpha_s}{4\pi}\Bigg\{
4\left(C_F-\frac{C_A}{2}\right) C_{24}(\{1\})
 +C_A \Bigg[ -\frac{1}{2}+2C_{24}(\{2\})-m_g^2C_0(\{2\})
    \nonumber \\ & & 
 +s \left(C_{23}(\{2\})+C_{12}(\{2\}) \right) 
 -\frac{1}{3}\left(\frac{2m_g^2}{s}+1\right)
\left(B_0(\{1\})+1\right) +\frac{2}{3s}A(m_g)+\frac{4}{9}
\Bigg] \nonumber \\ & &
 -\frac{1}{3}T_R\left[\left(1-\frac{4m_s^2}{s}\right)
\left(B_0(\{2\})+1\right) +\frac{4}{s}A(m_s)-\frac{1}{3}\right]
 \Bigg\}, 
\label{ff1} 
\end{eqnarray} 
where the Veltman-Passarino (VP) functions~\cite{VP}
$ C_i, \, B_0$ are given in the appendix. They are calculated in 
$n=4-2\epsilon$ dimensions, and the arguments $\{1\}, \, 
\{2\}$ are defined by
\begin{eqnarray}
C_i(\{1\})=C_i(s,m_s,m_g,m_s), \;
C_i(\{2\})=C_i(s,m_g,m_s,m_g), \nonumber \\
B_0(\{1\})=B_0(s,m_g,m_g), \;
B_0(\{2\})=B_0(s,m_s,m_s),
\label{arguments1}
\end{eqnarray}
where $m_s$ and $m_g$ are the squark and gluino masses, respectively.
For the unsubtracted correction to the form factor, we must add
to (\ref{ff1}) the pole term 
\begin{equation}
-C_F \frac{\alpha_s}{\epsilon},
\label{poleterm}
\end{equation}
which arises from the wave-function renormalization of the external quarks,
and cancels the pole part of (\ref{ff1})
proportional to $C_F$, as required by the abelian
Ward identity. Substituting the ultraviolet-divergent functions by their 
pole parts,
given in (\ref{pol0}-\ref{pol2}), we see that the poles from the one-particle
irreducible vertex corrections shown in Figs.~\ref{axofig}(a,b)) which are
proportional to $C_A$ also cancel between the two diagrams. The
the remaining divergences come from the functions $B_0$, which contribute
to the extra renormalization of the strong coupling 
due to supersymmetric particles, namely
 \begin{equation} 
-\frac{\alpha_s}{12\pi \epsilon}\left( C_A+T_R\right),
 \label{deltag} \end{equation}
and arise from the gluon self-energy contributions shown in 
Figs.~\ref{axofig}(g,h)).

Once this renormalization has been effected, the finite form factor is given
by (\ref{ff1}), with  the divergent functions $B_0$ and $C_{24}$ 
understood to have been subtracted 
using the $\overline{MS}$ prescription. 
The resulting contribution to the form factor
vanishes in the limit $s \to 0$
when we set $m_s=m_g=m$.  In the limit
$s \ll m^2$ the form-factor correction is given by the 
following simple expression:
\begin{equation}
 \lim_{s \ll m^2} \Delta F_{qqg} \to \frac{\alpha_s}{4\pi}\left\{ \frac{C_F}{12}+\frac{C_A}{60}
 -\frac{T_R}{30}\right\} \frac{s}{m^2}. \label{lowe1} \end{equation}
which is ${\cal O}(0.1 (s / m^2)) \%$. This and other form-factor 
corrections are
actually numerically considerably larger when $s \sim m^2$~\cite{thin},
as we shall see later.
\bigskip

The amplitude for the process (\ref{qqgg}), again omitting for the
moment box diagrams, may be written as
\begin{eqnarray}
- i 4\pi\alpha_s \epsilon_3^\nu\epsilon_4^\rho &  & \hspace*{-.1in} \Bigg\{
\frac{1}{t}\bar{v}(p_2)\tau^{a_4}\tau^{a_3}F_{gqq}^\rho(p_1-p_3)
 (p_1 \slash - p_3 \slash) F_{gqq}^\nu(p_1-p_3) u(p_1) \nonumber   \\ & &
+ \frac{1}{u}\bar{v}(p_2)\tau^{a_3}\tau^{a_4}F_{gqq}^\nu(p_1-p_4)
 (p_1 \slash - p_4 \slash) F_{gqq}^\rho(p_1-p_4) u(p_1) \nonumber \\ & &
-\frac{1}{s}\bar{v}(p_2)\left[\tau^{a_3},\tau^{a_4}\right]F_{qqg}(s)\gamma^\mu
 V_3^{\mu\nu\rho}(p_3,p_4)u(p_1)
  \Bigg\},
 \label{aqqgg} \end{eqnarray}
where $\epsilon_3, \, \epsilon_4$ are the polarization vectors of the external
gluons with momenta $p_3, \, p_4$, whose colours are $a_3,\, a_4$,
respectively.

The quantity $F_{gqq}^\mu(q)$ is the form factor for the 
gluon-quark-virtual-quark
vertex, which at the tree level is simply $\gamma^\mu$. The diagrams
contributing to the corrections to this form factor which involve
internal  squarks and gluinos are shown in Fig.~\ref{axofig}(e,f,i), and
give a contribution
\begin{eqnarray}
\Delta F_{gqq}^\mu (q) &= &\frac{\alpha_s}{4\pi} \Bigg\{
\left(4C_F-2C_A\right)\left[ C_{24}(\{4\})\gamma^\mu+
(C_{22}(\{4\})-C_{23}(\{4\})q \ss q^\mu \right] 
 \nonumber \\ & & \hspace*{-1in}
 +C_A\left[\left( 2 C_{24}(\{5\})-\frac{1}{2}-m_g^2C_0(\{5\})
 -q^2(C_{22}(\{5\})+C_{23}(\{5\})) \right)\gamma^\mu+2C_{23}(\{5\})q \ss
 q^\mu \right]
\nonumber \\ & & \hspace*{-.2in} -\frac{C_F}{2} \left[\frac{(m_g^2-m_s^2+q^2)}{q^2} B_0(q^2,m_s,m_g)
 -\frac{1}{q^2}\left(A(m_g)-A(m_s)\right)\right]\gamma^\mu \Bigg\},
 \label{ff2} \end{eqnarray}
where the arguments $\{4\}, \, \{5\}$ are given by
\begin{equation} C_i(\{4\})=C_i(q^2,m_s,m_s,m_g), \ 
 C_i(\{5\})=C_i(q^2,m_g,m_g,m_s). \label{arguments2} \end{equation}
To obtain the unsubtracted form factor, we must add a pole term
\begin{equation}
-\frac{\alpha_s}{4\pi\epsilon}\left(\frac{C_F}{2}+\frac{C_A}{3}
+\frac{T_R}{3} \right),
\label{poleterm2}
\end{equation}
which arises from the wave-function renormalizations of the external quark 
and gluon legs. Once this is included, 
we notice that again the abelian 
pole term proportional to $C_F$ cancels. 
The remaining pole term is given by (\ref{deltag}),
and is absorbed by the renormalization of the strong coupling. After this
renormalization, and setting $m_s=m_g=m$, we find that for $q^2 \ll m^2$
\begin{equation} \lim_{q^2 \ll m^2} \Delta F_{gqq}^\mu(q) \to
 \frac{\alpha_s}{4\pi}\left\{ C_A \left[ \frac{q^2}{4m^2} \gamma^\mu-
\frac{1}{3m^2} q \ss q^\mu\right] + \frac{C_F}{6m^2} q \ss q^\mu \right\} 
\label{lowe2} \end{equation}
which is ${\cal O}(1 (q^2 /m^2) \%$.

The three-gluon vertex function $V_3^{\mu\nu\rho}(p3,p4)$ may be written as
\begin{equation}
V_3^{\mu\nu\rho}(p_3,p_4)  =
 -2g^{\nu\rho}p_4^\mu F_1(s)+
 2g^{\mu\rho}p_4^\nu F _2(s)  
 -2g^{\mu\nu}p_3^\rho F_2(s)  
 +p_4^\nu p_3^\rho p_4^\mu F_3(s) \label{v3}, \end{equation}
where $s=2p_3 \cdot p_4$. At the tree level, the form factors $F_1, \, F_2$
take the value unity, whereas $F_3$ is zero. The 
irreducible one-loop contributions
from squark and gluino loops to these form factors are 
are shown in Fig.~\ref{axofig}(c,d), and are given by
\begin{eqnarray}
\Delta F_1(s)&=& \frac{\alpha_s}{4\pi}\Bigg\{ C_A\Bigg[
-\frac{2}{3}-m_g^2\left(C_0(\{6\})-C_{11}(\{6\})+C_{12}(\{6\})\right)
\nonumber \\ & & 
+s\left(C_{12}(\{6\})+C_{22}(\{6\})-C_{33}(\{6\})+C_{34}(\{6\})\right)
\nonumber \\ & &
-2C_{35}(\{6\})+2C_{36}(\{6\})+2C_{24}(\{6\}) \nonumber \\ & &
 -\frac{1}{3}
\left(\frac{2m_g^2}{s}+1\right) 
\left(B_0(\{1\})+1\right) +\frac{2}{3s}A(m_g)+\frac{4}{9}\Bigg] 
\nonumber \\ & & \hspace*{-1.1in}
 +T_R\left[8C_{36}(\{7\})-8C_{35}(\{7\})
-\frac{1}{3}\left(1-\frac{4m_s^2}{s}\right)
\left(B_0(\{2\})+1\right)-\frac{4}{3s}A(m_s)+\frac{1}{9}\right]
 \Bigg\}, \label{f1} \end{eqnarray}
\begin{eqnarray}
\Delta F_2(s)&=&\frac{\alpha_s}{4\pi}\Bigg\{ C_A \Bigg[
-\frac{2}{3}-m_g^2\left(C_0(\{6\})-C_{12}(\{6\})\right)
\nonumber \\ & & 
-s\left(C_{22}(\{6\})+C_{34}(\{6\})\right)
 -2C_{36}(\{6\})+2C_{24}(\{6\})  \nonumber \\ & &
 -\frac{1}{3}\left(\frac{2m_g^2}{s}+1\right)  
\left(B_0(\{1\})+1\right)+\frac{2}{3s}A(m_g)+\frac{4}{9}\Bigg] 
\nonumber \\ & & \hspace*{-.4in}
 +T_R\left[-8C_{36}(\{7\})
-\frac{1}{3}\left(1-\frac{4m_s^2}{s}\right)
\left(B_0(\{2\})+1\right)-\frac{4}{3s}A(m_s)+\frac{1}{9}\right]
 \Bigg\}, \label{f2} \end{eqnarray}
\begin{eqnarray}
\Delta F_3(s)&=&\frac{2 \alpha_s}{\pi}\Bigg\{C_A\left[
C_{22}(\{6\})-C_{23}(\{6\})-C_{33}(\{6\})+C_{34}(\{6\})\right] \nonumber \\ & &
+2 T_R\left[C_{23}(\{7\})-C_{22}(\{7\})+C_{33}(\{7\})-C_{34}(\{7\})
\right]\Bigg\}, \label{f3} \end{eqnarray}
where the arguments $\{6\}, \, \{7\}$ are given by
\begin{equation} C_i(\{6\})=C_i(s,m_g,m_g,m_g), \ 
 C_i(\{7\})=C_i(s,m_s,m_s,m_s). \label{arguments3} \end{equation}
To obtain the unsubtracted form factor, we must add a pole term
\begin{equation}
-\frac{\alpha_s}{4\pi\epsilon}\left(\frac{2C_A}{3}
+\frac{2T_R}{3} \right),
\label{poleterm3}
\end{equation}
which arises from the wave-function renormalizations of the external  
gluon legs as shown in Fig.~\ref{axofig}(g,h), leaving a pole term
equal to (\ref{deltag}).  After renormalization
and  setting $m_s=m_g=m$, we find that for $s \ll m^2$
\begin{equation}  \lim_{s \ll m^2} \Delta F_{1}(s) \to
 \frac{\alpha_s}{48\pi}  C_A \frac{s}{m^2}  \label{lowe3} \end{equation}
\begin{equation} \lim_{s \ll m^2} \Delta F_{2}(s) \to
 \frac{\alpha_s}{60\pi}  \left(C_A+\frac{T_R}{2}\right)
  \frac{s}{m^2}  \label{lowe4} \end{equation}
 \begin{equation} \lim_{s \ll m^2} \Delta F_{3}(s) \to
 \frac{\alpha_s}{60\pi} \left( C_A-2T_R \right) \frac{1}{m^2}  \label{lowe5} \end{equation}
which are comparable in magnitude to the low-energy expansions
of the previous form factors. Apart from
some minor differences, the low-energy
expansions of our results 
(eqs.(\ref{lowe1},\ref{lowe2}, \ref{lowe3}, \ref{lowe4}, \ref{lowe5}))
confirm the general magnitudes of the below-threshold corrections
found in \cite{KW}.
\bigskip

Finally, we can use $V_3$ to write down the amplitude for process 
(\ref{gggg})
excluding the box diagrams which contribute to the renormalization of the
four-gluon coupling, which we postpone until the next section:
\begin{equation}
i\frac{4\pi \alpha_s}{s}
\epsilon_1^\mu\epsilon_2^\nu\epsilon_3^\rho\epsilon_4^\sigma
f^{a_1a_2b}f^{a_3a_4b} V_3^{\tau\mu\nu}(p_1,p_2)
V_3^{\tau\rho\sigma}(p_3,p_4)  \ + \ {\rm permutations} \ \{2,3,4\}, 
\label{agggg} \end{equation}
where $\epsilon_1, \, \epsilon_2$ are the polarization vectors of the incoming
gluons with momenta $p_1, \, p_2$, whose colours are $a_1,\, a_2$,
respectively,
and $\epsilon_3, \, \epsilon_4$ are the polarization vectors of the outgoing
gluons with momenta $p_3, \, p_4$, whose colours are $a_3,\, a_4$,
respectively. The notation ``permutations $\{2,3,4\}$'' means permutations
over the momenta, polarization vectors and colours of gluons 2,3,4.

\section{Box Diagrams}

For the box diagrams we need the four-point
VP functions $D_i$~\cite{VP} listed in the appendix in
(\ref{four1}) to (\ref{four5}).

\begin{figure}
\begin{center}
\leavevmode
\hbox{\epsfxsize=4.0 in
\epsfbox{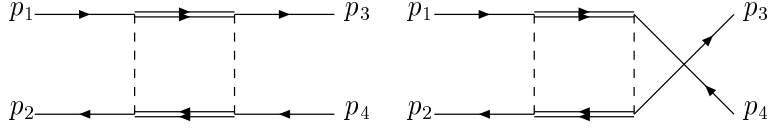}}
\end{center}
\vspace*{-.2in}
\caption[]{Box diagrams for the process $q \, \bar{q} \to q \, \bar{q}$.} 
\label{qqqqbox}
\end{figure}

The box diagrams for the process (\ref{qqqq}),
shown in Fig. \ref{qqqqbox},  give the following contribution to the 
amplitude:
\begin{eqnarray}
 2 i C_A\alpha_s^2 \left\{  
D_2(t,u,m_s,m_g,m_s,m_g,\mu,\nu) \left[ A_{VV}^{\mu\nu}+A_{AA}^{\mu\nu} \right]
 \right. & & \nonumber \\  \left. +
m_g^2 D_0(t,u,m_s,m_g,m_s,m_g)\left[ A_{SS}+A_{PP} \right] \right\} &+ \; (t 
\leftrightarrow u) & 
 \end{eqnarray}
where
$$A_{VV}^{\mu\nu}=\bar{v}(p_4)\gamma^\mu \tau^a v(p_2) \ \bar{u}(p_3)\gamma^\nu
 \tau^a u(p_1)$$
$$A_{AA}^{\mu\nu}=\bar{v}(p_4)\gamma^\mu \gamma^5 \tau^a v(p_2) \ 
 \bar{u}(p_3)\gamma^\nu \gamma^5
 \tau^a u(p_1)$$
$$A_{SS}=\bar{v}(p_4) \tau^a v(p_2)  \ \bar{u}(p_3)\tau^a u(p_1)$$
$$A_{PP}=\bar{v}(p_4)\gamma^5 \tau^a v(p_2) \  \bar{u}(p_3)\gamma^5
 \tau^a u(p_1)$$
The second graph in Fig. \ref{qqqqbox} arises from the Majorana nature of the
internal gluino lines, which permits them to propagate both along
and against the flow of the quark fermion number. 
\bigskip

\begin{figure}
\begin{center}
\leavevmode
\hbox{\epsfxsize=6.0 in
\epsfbox{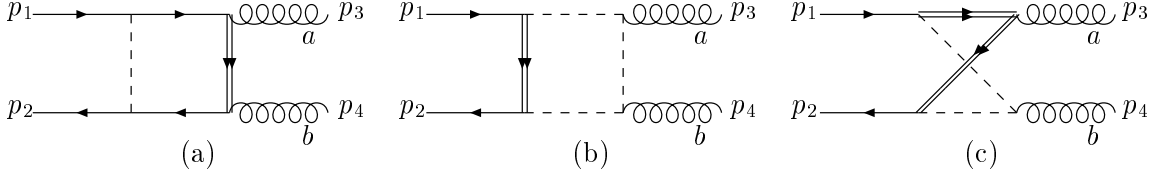}}
\end{center}
\vspace*{-.3in}
\caption[]{Box diagrams for the process $q \, \bar{q} \to g \, g$.}
 \label{qqggbox}
\end{figure}

The box diagrams for the process 
(\ref{qqgg})
are shown in Fig. \ref{qqggbox}.
Fig. \ref{qqggbox}(a) gives a contribution
\begin{eqnarray}
-2i \alpha_s^2 \epsilon_3^\nu \epsilon_4^\sigma \left\{ \right. & & 
\hspace*{-.25in}
\left[ \, D_3(t,u,m_s,m_g,m_g,m_g,\mu,\rho,\tau)  \right. \nonumber \\ & &
 \hspace*{-.75in} \left.
+ \, (p_1-p_3)^\rho
D_2(t,u,m_s,m_g,m_g,m_g,\mu,\tau)\right] \, B_{ab}^{\mu\nu\rho\sigma\tau}
\nonumber \\ & & \hspace*{-1.25in} + \, m_g^2 \left[ \,
 D_1(t,u,m_s,m_g,m_g,m_g,\mu) \left(
B_{ab}^{\nu\mu\sigma}+B_{ab}^{\mu\nu\sigma}+B_{ab}^{\nu\sigma\mu} \right)
\right. \nonumber \\ & & \hspace*{-.25in} \left.
+ \, (p_1-p_3)^\mu D_0(t,u,m_s,m_g,m_g,m_g) B_{ab}^{\nu\mu\sigma} \right]
\left. \right\}
 \ + (3 \leftrightarrow 4),
\end{eqnarray}
where the tensors $B$ are defined as
\begin{equation}
B_{ab}^{\mu_1 \cdots \mu_n}=\bar{v}(p_2)\left( \gamma^{\mu_1} \cdots \gamma^{\mu_n}
\left[ \tau^b,\tau^c \right] \Big[ \tau^a, \tau^c \Big] \right)u(p_1).
\label{defineB}
\end{equation}
Fig. \ref{qqggbox}(b) gives a contribution
\begin{eqnarray}
-8i \alpha_s^2 \epsilon_3^\nu \epsilon_4^\sigma  & & 
\hspace*{-.25in}
\left\{ \, D_3(t,u,m_g,m_s,m_s,m_s,\mu,\nu,\sigma)
+p_1^\nu D_2(t,u,m_g,m_s,m_s,m_s,\mu,\sigma)
 \right. \nonumber \\ & &
 \hspace*{-.9in} + \, \left.
(p_1-p_3)^\sigma \left[ \, D_2(t,u,m_g,m_s,m_s,m_s,\mu,\nu )
 +  p_1^\nu  D_1(t,u,m_g,m_s,m_s,m_s,\mu) \right] \right\}
\nonumber \\ & & \hspace*{-.25in}
 \bar{v}(p_2)\left( \gamma^\mu \tau^c \tau^b \tau^a \tau^c \right) u(p_1)
 \ + (3 \leftrightarrow 4)
\end{eqnarray}
Fig. \ref{qqggbox}(c) gives a contribution
\begin{eqnarray}
-4i \alpha_s^2 \epsilon_3^\nu \epsilon_4^\sigma \left\{ \right. & & 
\hspace*{-.25in}
\left[ \, D_3(t,s,m_s,m_g,m_g,m_s,\mu,\rho,\sigma)  \right. \nonumber \\ & &
 \hspace*{-1.in} \left.
+ \, (p_1-p_3)^\mu
D_2(t,s,m_s,m_g,m_g,m_s,\rho,\sigma)\right] \, B_{ab}^{\prime\mu\nu\rho}
\nonumber \\ & & \hspace*{-.25in} + \, m_g^2  \,
 D_1(t,s,m_s,m_g,m_g,m_s,\sigma) B_{ab}^{\prime \nu} \left. \right\}
 \ + (3 \leftrightarrow 4),
\end{eqnarray}
where the tensors $B^\prime$ are defined by
\begin{equation}
B_{ab}^{\prime\mu_1 \cdots \mu_n}=if^{dac} \, \bar{v}(p_2)\left( \gamma^{\mu_1}
 \cdots \gamma^{\mu_n}  \tau^c \tau^b \tau^d \right)u(p_1).
\label{defineBprime}
\end{equation}
and the notation $(3 \leftrightarrow 4)$ means
$(\epsilon_3 \leftrightarrow \epsilon_4, \  p_3 \leftrightarrow p_4,
 \ t \leftrightarrow u, \ a \leftrightarrow b )$.
\bigskip

\begin{figure}
\begin{center}
\leavevmode
\hbox{\epsfxsize=2.0 in
\epsfbox{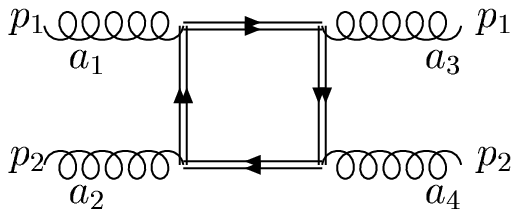}}
\end{center}
\vspace*{-.2in}
\caption[]{Gluino box diagram for the process $g \, g \to g \, g$.}
\label{ggggbox1}
\end{figure}

\begin{figure}
\begin{center}
\leavevmode
\hbox{\epsfxsize=4.0 in
\epsfbox{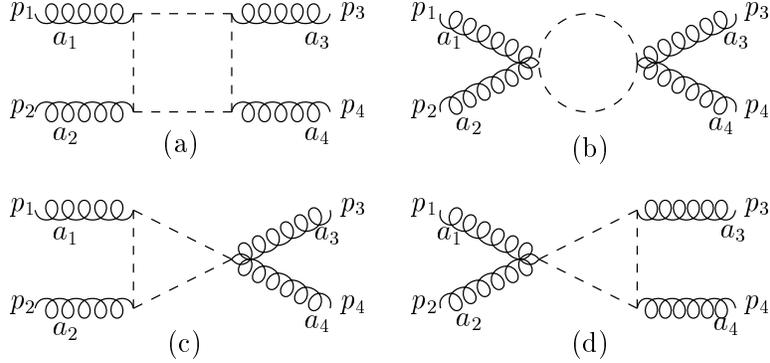}}
\end{center}
\vspace*{-.2in}
\caption[]{(a) Squark box diagram and (b,c,d) diagrams involving quartic 
couplings that contribute to the process $g \, g \to g \, g$.}
\label{ggggbox2}
\end{figure}

The contributions to the process (\ref{gggg})
from box diagrams are shown in Fig. \ref{ggggbox1} (for an internal gluino 
loop) and Fig. \ref{ggggbox2} (for an internal squark loop).
Also included in Fig. \ref{ggggbox2} are diagrams that
involve the four-point coupling between gluons and squarks. 
Although these are not box diagrams, we include them here,
because they do not contribute to the form factor
for the triple-gluon vertex.
The contribution to the amplitude from Fig. \ref{ggggbox1} is
\begin{eqnarray}
i \frac{\alpha_s^2}{2} G_1 & & \hspace*{-.2in} \left\{ \right.
D_4(s,t,m_g,m_g,m_g,m_g,\mu,\nu,\rho,\sigma){\rm tr}
\left(\epsilon_1 \slash \gamma^\mu \epsilon_4 \slash \gamma^\nu
 \epsilon_3 \slash \gamma^\rho \epsilon_2 \slash \gamma^\sigma \right)
 \nonumber \\ & & \hspace*{-.3in}
+D_3(s,t,m_g,m_g,m_g,m_g,\mu,\nu,\rho) \left[ {\rm tr}
\left(\epsilon_1 \slash \gamma^\mu \epsilon_4 \slash p_4 \slash
 \epsilon_3 \slash \gamma^\nu \epsilon_2 \slash \gamma^\rho \right) \right.
\nonumber \\ & & 
+ {\rm tr}
\left(\epsilon_1 \slash \gamma^\mu \epsilon_4 \slash \gamma^\nu
 \epsilon_3 \slash \gamma^\rho \epsilon_2 \slash p_1 \slash \right) + {\rm tr}
\left(\epsilon_1 \slash \gamma^\mu \epsilon_4 \slash \gamma^\nu
 \epsilon_3 \slash (p_1 \slash + p_2 \slash) \epsilon_2 \slash \gamma^\rho \right)
\left. \right] \nonumber \\ & & \hspace*{-.3in}
+D_2(s,t,m_g,m_g,m_g,m_g,\mu,\nu) \left[ {\rm tr}
\left(\epsilon_1 \slash \gamma^\mu \epsilon_4 \slash p_4 \slash
 \epsilon_3 \slash \gamma^\nu \epsilon_2 \slash p_1 \slash \right) \right.
\nonumber \\ & &
+ {\rm tr}
\left(\epsilon_1 \slash \gamma^\mu \epsilon_4 \slash p_4 \slash
 \epsilon_3 \slash (p_1 \slash + p_2 \slash)
 \epsilon_2 \slash \gamma^\nu  \right) + {\rm tr}
\left(\epsilon_1 \slash \gamma^\mu \epsilon_4 \slash \gamma^\nu
 \epsilon_3 \slash (p_1 \slash + p_2 \slash) \epsilon_2 \slash p_1 \slash \right)
\left. \right] \nonumber \\ & & \hspace*{-.3in}
+D_1(s,t,m_g,m_g,m_g,m_g,\mu)  {\rm tr} \left(
\epsilon_1 \slash \gamma^\mu \epsilon_4 \slash p_4 \slash \epsilon_3 \slash
 \left( p_1 \slash + p_2 \slash \right) \epsilon_2 \slash p_1 \slash \right)
 \nonumber \\ & & \hspace*{-.5in}
+m_g^2 D_2(s,t,m_g,m_g,m_g,m_g,\mu,\nu)
 \left[ 16 \epsilon_1^\mu \epsilon_3^\nu  \epsilon_2 \cdot \epsilon_4
 +   16 \epsilon_4^\mu \epsilon_2^\nu  \epsilon_1 \cdot \epsilon_3    \right]
\nonumber \\ & & \hspace*{-.5in}
+m_g^2 D_1(s,t,m_g,m_g,m_g,m_g,\mu) \left[ 2 \epsilon_4^\mu {\rm tr}\left(
 \epsilon_1 \slash \epsilon_3 \slash \left( p_1 \slash + p_2 \slash \right)
\epsilon_2 \slash \right) \right. \nonumber \\ & & \hspace*{-.2in}
+ 2 \epsilon_2^\mu {\rm tr} \left( \epsilon_1 \slash  \epsilon_4 \slash
 p_4 \slash \epsilon_3 \slash \right) + 2 \epsilon_3^\mu {\rm tr}
 \left( \epsilon_1 \slash \epsilon_4 \slash \epsilon_2 \slash p_1 \slash \right)
+ {\rm tr} \left( \epsilon_1 \slash \gamma^\mu \epsilon_4 \slash p_4 \slash
\epsilon_3 \slash \epsilon_2 \slash \right)
\nonumber \\ & & 
+ {\rm tr} \left( \epsilon_1 \slash \gamma^\mu \epsilon_4  \slash
\epsilon_3 \slash \epsilon_2 \slash p_1 \slash \right) 
 + {\rm tr} \left( \gamma^\mu \epsilon_1 \slash  \epsilon_4  \slash
\epsilon_3 \slash \left(p_1 \slash + p_2 \slash \right)
\epsilon_2 \slash \right) \left. \right]
\nonumber \\ & & \hspace*{-.5 in}
+m_g^2 D_0(s,t,m_g,m_g,m_g,m_g)\left[ {\rm tr}
\left( \epsilon_1 \slash \epsilon_4 \slash p_4 \slash \epsilon_3 \slash
 \left( p_1 \slash + p_2 \slash \right) \epsilon_2 \slash \right) \right.
 \nonumber \\ & & 
 + {\rm tr} \left( \epsilon_1 \slash \epsilon_4 \slash \epsilon_3 \slash 
\left( p_1 \slash + p_2 \slash \right) \epsilon_2 \slash p_1 \slash \right)
 + {\rm tr} \left( \epsilon_1 \slash \epsilon_4  p_4 \slash\slash \epsilon_3 \slash 
    \epsilon_2 \slash p_1 \slash \right)
\left. \right] 
\nonumber \\ & & \hspace*{-.5in}
+m_g^4 D_0(s,t,m_g,m_g,m_g,m_g){\rm tr} \left( \epsilon_1 \slash
 \epsilon_4 \slash \epsilon_3 \slash \epsilon_2 \slash \right)
\left. \right\} 
\nonumber \\ & & (+ \ {\rm permutations} \ \{ 2,3,4\}),
\label{ghastly}
\end{eqnarray}  
where the colour factor $G_1$ is given by
\begin{equation}
G_1= f^{a_1ab}f^{a_2bc}f^{a_3cd}f^{a_4da} . 
\label{defineG1} 
\end{equation}
In the interests of compactness,
we have left this expression in terms of traces over $\gamma$ matrices.
The contribution from Fig. \ref{ggggbox2}(a) is
\begin{eqnarray}
- 32 i\alpha_s^2 G_2 \, \epsilon_1^\mu \epsilon_2^\nu \epsilon_3^\rho \epsilon_4^\sigma 
  & & \hspace*{-.2in} 
 \left\{ \right.   D_4(s,t,m_s,m_s,m_s,m_s,\mu,\nu,\rho,\sigma)
\nonumber \\ & & \hspace*{-.2in}
+p_1^\nu D_3(s,t,m_s,m_s,m_s,m_s,\mu,\rho,\sigma)
\nonumber \\ & & \hspace*{-.2in}
+\left( p_1+p_2 \right)^\rho  D_3(s,t,m_s,m_s,m_s,m_s,\mu,\nu,\sigma)
\nonumber \\ & & \hspace*{-.2in}
p_1^\nu (p_1+p_2)^\rho D_2(s,t,m_s,m_s,m_s,m_s,\mu,\sigma) \left. \right\}
\nonumber \\ & & 
 \ + \ {\rm permutations} \ \{ 2,3,4 \} ), 
\label{lessghastly}
\end{eqnarray}
where the colour factor $G_2$ is given by
\begin{equation} 
G_2= {\rm tr}\left(\tau^{a_1}\tau^{a_2}\tau^{a_3}\tau^{a_4} \right)
\label{G2} 
\end{equation}
The contribution from Fig. \ref{ggggbox2}(b) is
\begin{eqnarray} & & \hspace*{-.2in} 
4 i \alpha_s^2 G_3  \, \epsilon_1 \cdot \epsilon_4 \, \epsilon_2 \cdot \epsilon_3 \, 
  \, B_0(u,m_s,m_s) \\ & & 
\  (+ \ {\rm permutations} \ \{ 2,3,4\}),
\label{partc}
\end{eqnarray}
where the colour factor $G_3$ is given by
\begin{equation}
G_3= {\rm tr}\left(\tau^{a_1}\tau^{a_2}\tau^{\{a_3}\tau^{a_4\}} \right)
\label{G3} 
\end{equation}

The contribution from Fig, \ref{ggggbox2}(c) is
\begin{eqnarray} & & \hspace*{-.2in}
- 16 i \alpha_s^2 G_3 \,\epsilon_2 \cdot \epsilon_3  \, \left[
  \, C_{24}(u,m_s,m_s,m_s) \, \epsilon_1 \cdot \epsilon_4
   - C_{23}(u,m_s,m_s,m_s) \, p_4 \cdot \epsilon_1 \,  p_1 \cdot
 \epsilon_4 \right] \nonumber \\ & & 
 \  (+ \ {\rm permutations} \ \{ 2,3,4\}),
\label{partc2}
\end{eqnarray}
Finally, from Fig. \ref{ggggbox2}(d) we get
\begin{eqnarray} & & \hspace*{-.2in}
- 16 i \alpha_s^2 G_3 \, \epsilon_1 \cdot \epsilon_4  \, \left[
  \, C_{24}(u,m_s,m_s,m_s) \, \epsilon_2 \cdot \epsilon_3
   - C_{23}(u,m_s,m_s,m_s) \, p_3 \cdot \epsilon_2 \,  p_2 \cdot
 \epsilon_3 \right] \nonumber \\ & & 
 \  (+ \ {\rm permutations} \ \{ 2,3,4 \} ),
\label{partd}
\end{eqnarray}
These contributions contain ultraviolet divergences, which are associated
with the renormalization $Z_4$ of the four-point gluon vertex. Application of
Ward identities tells us that the coefficient of the 
pole parts of the integrals should be given by the
contribution $\Delta \beta_0$ to the $\beta$ function, multiplied by the 
tree-level amplitude $\Gamma_4$ for the
four-gluon  coupling. In the above expressions, 
poles arise in the unsubtracted forms of the functions
$B_0, \, C_{24}\, D_4$.
Exploiting the pole parts displayed in the appendix
in (\ref{pol1},\ref{pol2},\ref{pol5}),
and permuting the gluons 2,3,4, we arrive
at a pole term
\begin{equation}
-\frac{\alpha_s}{4\pi\epsilon} \frac{2}{3} \left( C_A+T_R \right) 
\Gamma_4 = \frac{1}{\epsilon}\Delta \beta_0 \Gamma_4, 
\label{check}
\end{equation}
as required.
\bigskip

In order to obtain the relevant contributions
to the partonic cross sections, these contributions to amplitudes
must be multiplied by the Hermitian conjugates of the corresponding
tree-level amplitudes and summed (averaged) over final (initial)
quark or gluon polarizations and colours. These tedious but straightforward
manipulations are most conveniently performed using a fast algebraic
manipulation package: we have used FORM.

Finally, we note that the box-diagram contributions for all other partonic
subprocesses can be obtained from the above expressions by crossing symmetry.

\section{Numerical Results}

\begin{figure}
\hbox to \hsize{\hss
\epsfxsize=0.49\hsize
\epsffile{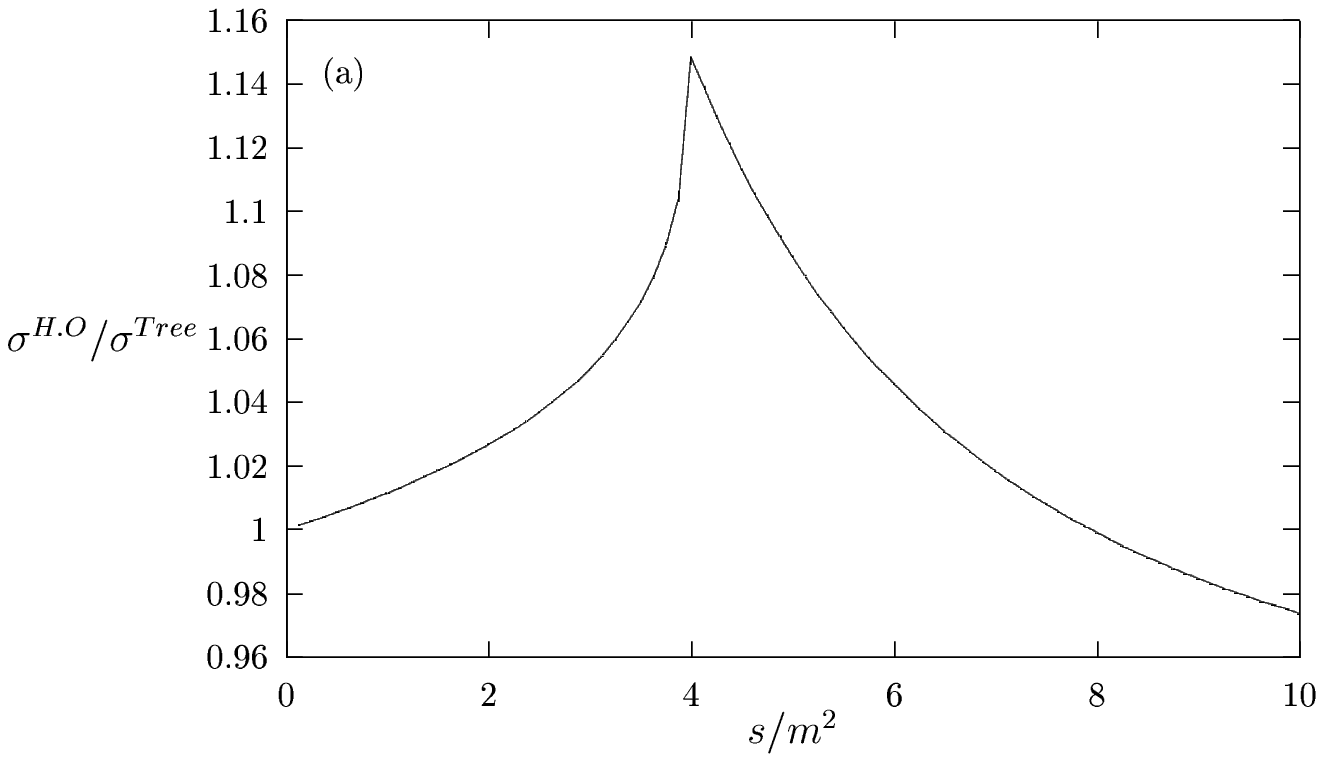}
\hfill
\epsfxsize=0.49\hsize
\epsffile{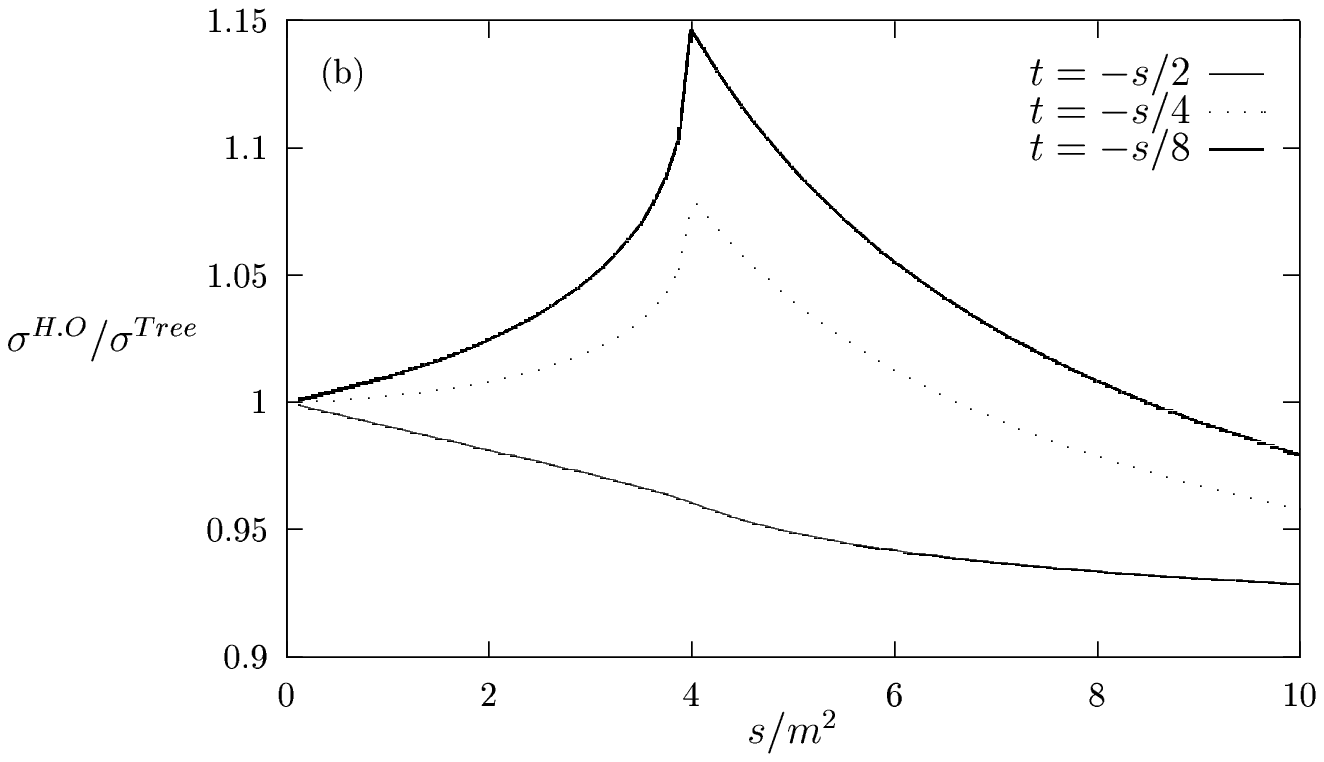}
\hss}
\kern2em
\hbox to \hsize{\hss
\epsfxsize=0.49\hsize
\epsffile{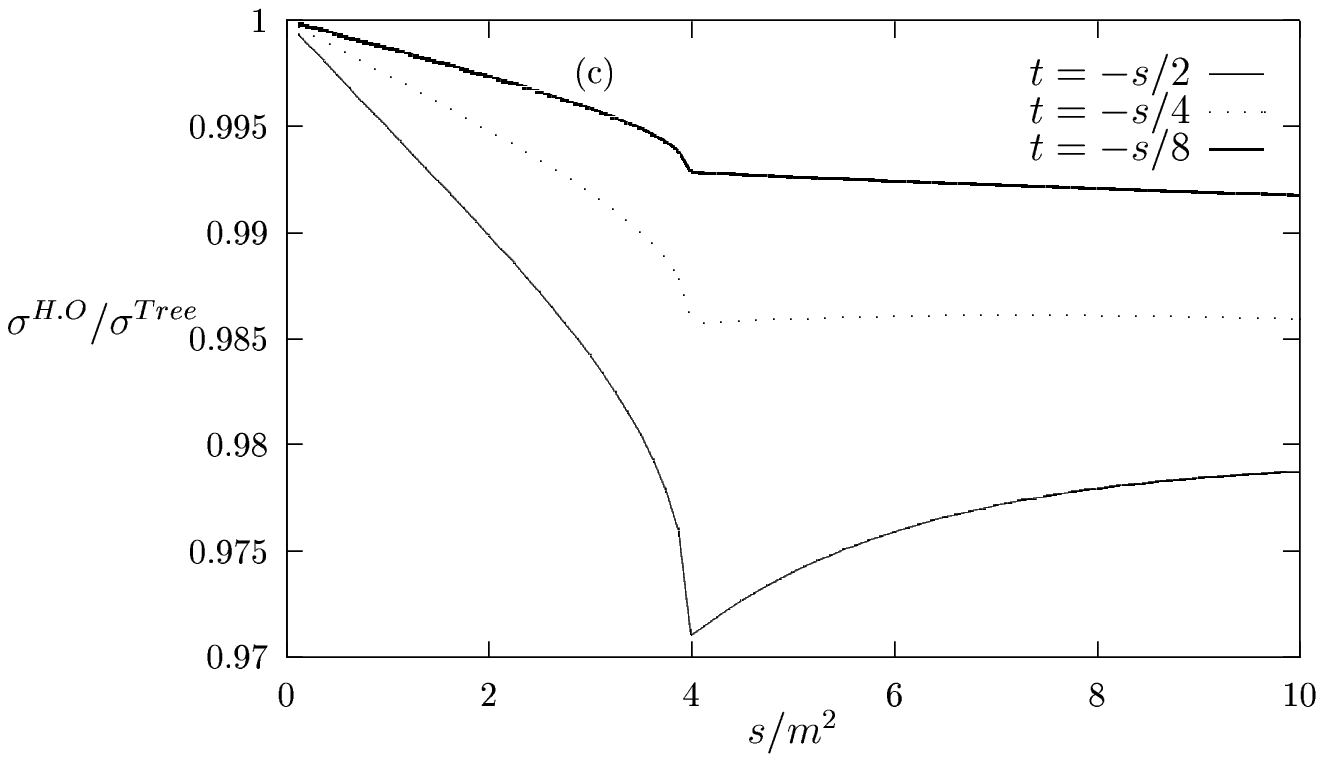}
\hfill
\epsfxsize=0.49\hsize
\epsffile{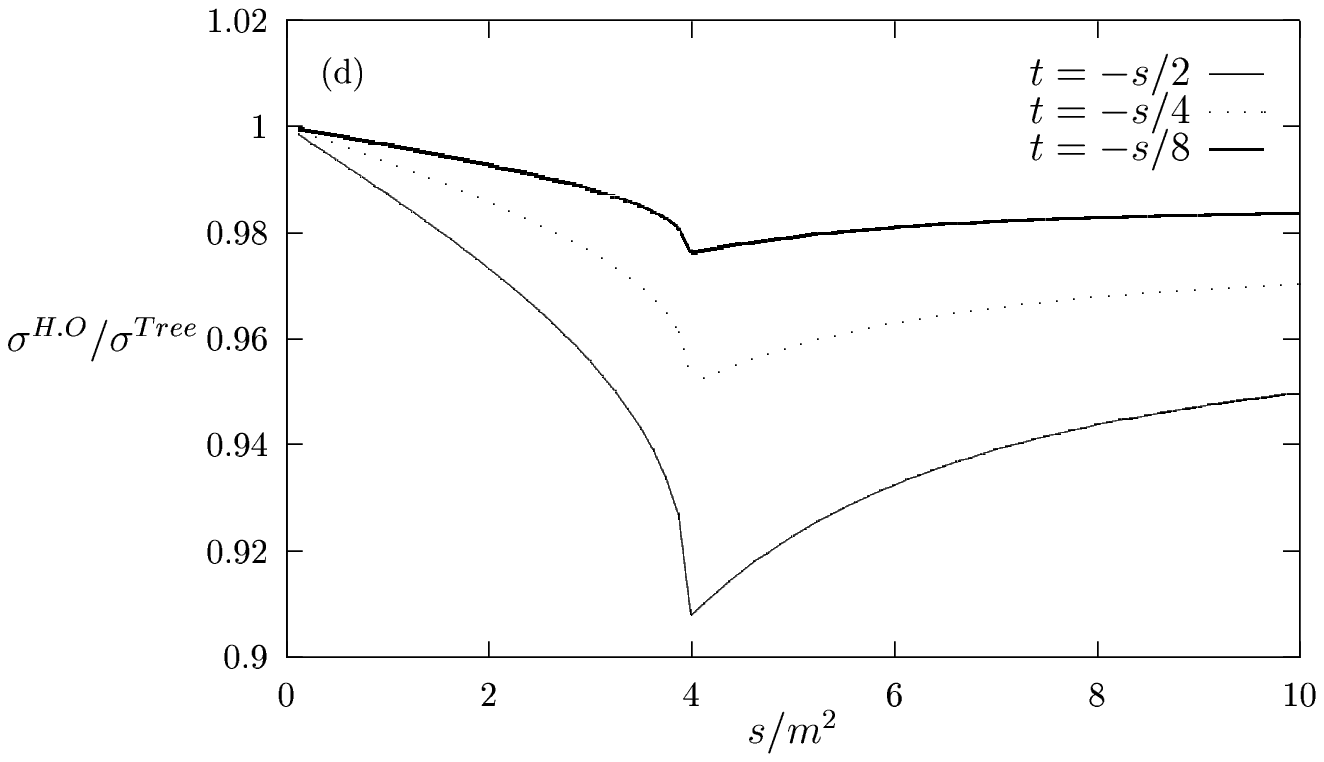}
\hss}
\kern2em
\hbox to \hsize{\hss
\epsfxsize=0.49\hsize
\epsffile{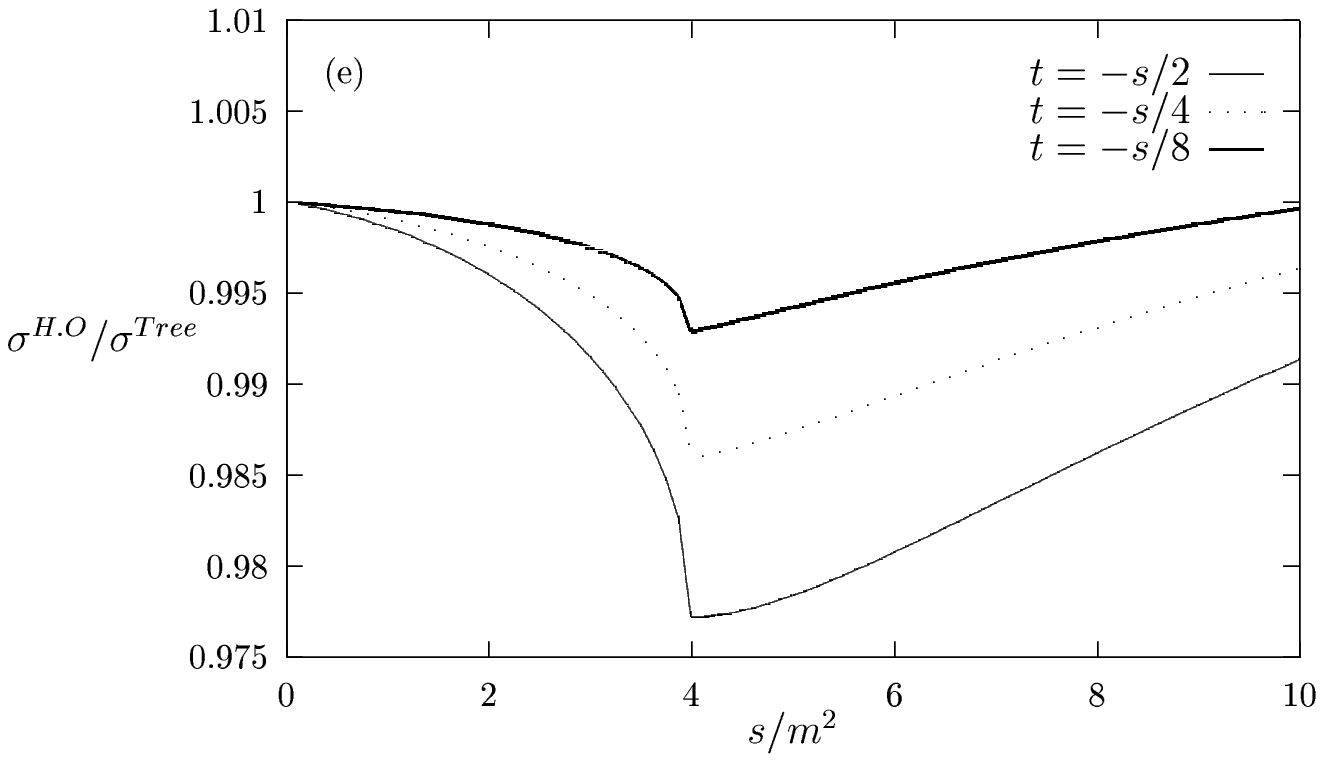} \hss}
\caption[]{One-loop virtual-sparticle corrections in the threshold
region of the subprocess centre-of-mass energy squared $s$
to the processes (a)
$q_j \, \bar{q}_j \rightarrow q_k \, \bar{q}_k$,
 (b) $q  \, \bar{q} \rightarrow g \, g$, 
(c) $q_j \, q_k \rightarrow q_j \, q_k$,  (d) $q_j \, g \rightarrow
q_j \, g$, and (e) $g \, g \rightarrow g \, g$.
In each case we have plotted the ratio of the one-loop 
cross section  to the tree-level value, and in (b,c,d,e) we have 
plotted this ratio for three different choices of $t/s$, 
equivalent to different parton-parton centre-of-mass scattering
angles.  All corrections are evaluated
using $\alpha_s=0.11$, and $m_s=m_g=m$.}
\label{subprocesses}
\end{figure}

We first present the results for the one-loop corrections to all the 
different parton-parton scattering cross sections shown in 
Fig.~\ref{subprocesses}. These are for the simplified case $m_s = m_g = 
m$, although our analytical results apply for arbitrary ratios
$m_s / m_g$. These extend and complete the analogous plots shown
in~\cite{thin}, which did not include box diagrams, nor
results for the subprocess $gg \rightarrow gg$. We already
commented in~\cite{thin} that we did not expect the box-diagram
contributions to be numerically large, and this has been confirmed
by our complete calculation. 

We first 
note that, because these corrections to the cross sections are due to
interferences between tree-level and one-loop amplitudes, they are
not necessarily positive. In particular,
processes which are dominated by the exchange of a parton in the $t$ channel
give a negative correction, and
we see that the corrections in
Fig.~\ref{subprocesses}(c,d,e) are negative for $s \sim 4 m^2$. We also
note that each of the corrections exhibits a cusp at $\sqrt{s} = 2 m$,
creating a local maximum in the magnitude of the one-loop correction. 
This is largest
for the subprocesses $q_j {\bar q}_j \rightarrow q_k {\bar q}_k$ and
$q {\bar q} \rightarrow gg$ shown in Figs.~\ref{subprocesses}(a,b), which 
are unfortunately not dominant at
the Fermilab Tevatron collider and the LHC. Next in magnitude is the
$q_j g \rightarrow q_j g$ shown in Fig.~\ref{subprocesses}, which is
of greater experimental significance. The newly-calculated
one-loop correction to the
subprocess $g g \rightarrow g g$ shown in Fig.~\ref{subprocesses}(e)
is the smallest numerically.

We also note that each of the one-loop corrections grows logarithmically
at large $s \gg m^2$. As was mentioned previously, some of
this logarithmic behaviour is due to ultraviolet divergences associated 
with the different running of $\alpha_s$ above the sparticle threshold.
However, this is not the only source of logarithmic behaviour. Since
the two-jet cross section does not include final states 
containing sparticles, there are also infrared divergences when
$m^2 \ll s$. Moreover, numerical studies show that the 
logarithmic asymptotic behaviour does not set in until subenergies
beyond the reach of the Fermilab Tevatron collider - for $m = 200$ GeV -
or the LHC - for $m = 1$ TeV. For these two reasons, it is not
adequate to model the sparticle threshold simply by switching to
the above-threshold form of $\alpha_s$ for $s > 4 m^2$.

\begin{figure}
\hbox to \hsize{\hss
\epsfxsize=0.69\hsize
\epsffile{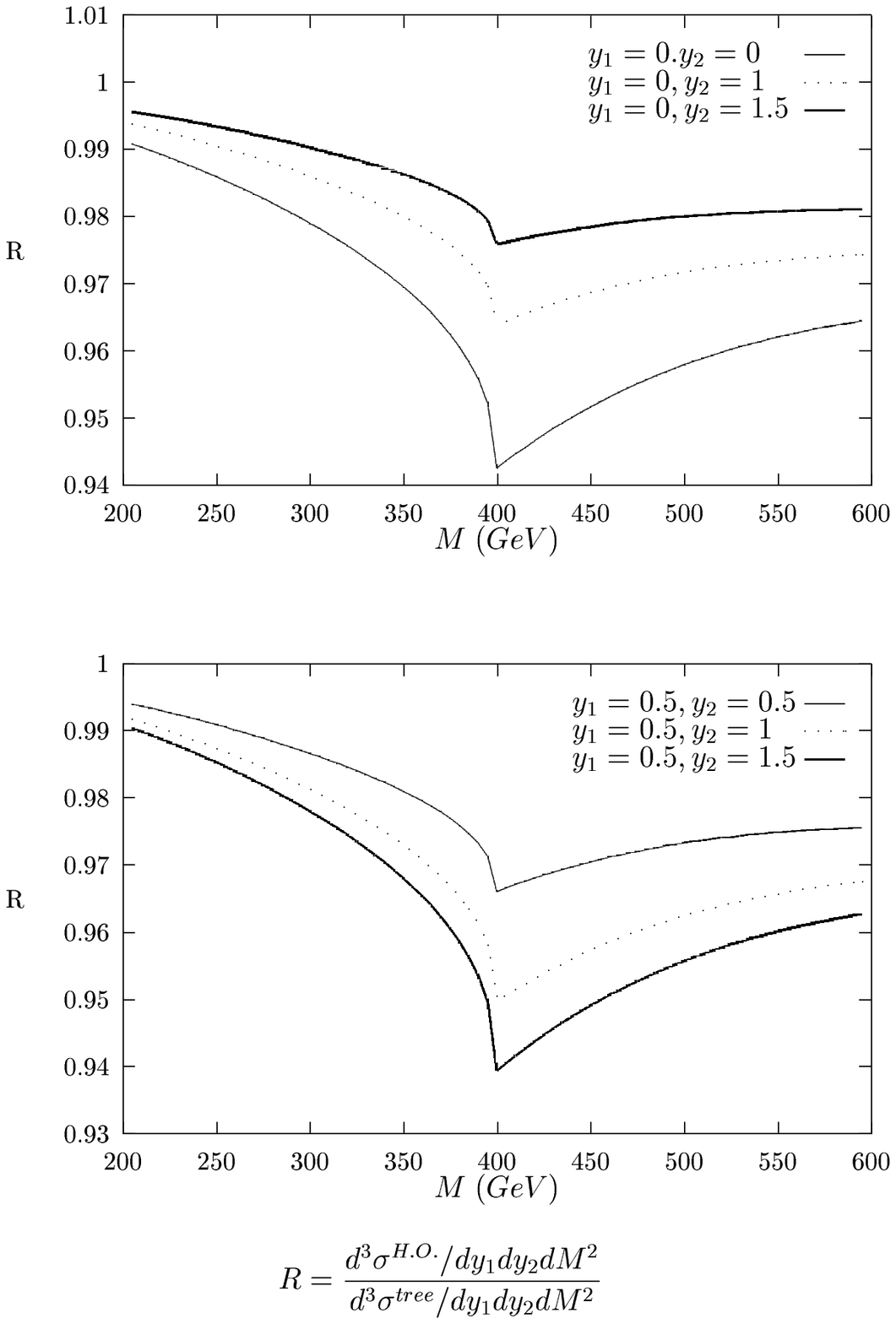}
\hss}
\caption[]{
Triple-differential cross section for $p \, \bar{p}$ scattering
at $\sqrt{s}=1.8$ TeV, as a function of the two-jet invariant mass $M$, for
various choices of jet rapidities, calculated assuming $m_s=m_g=200$ GeV. 
Here and in subsequent plots, $R$ denotes the ratio
of the one-sparticle-loop-corrected cross section to the tree-level
cross section.}
\label{tev1}
\end{figure}

\begin{figure}
\hbox to \hsize{\hss
\epsfxsize=0.69\hsize
\epsffile{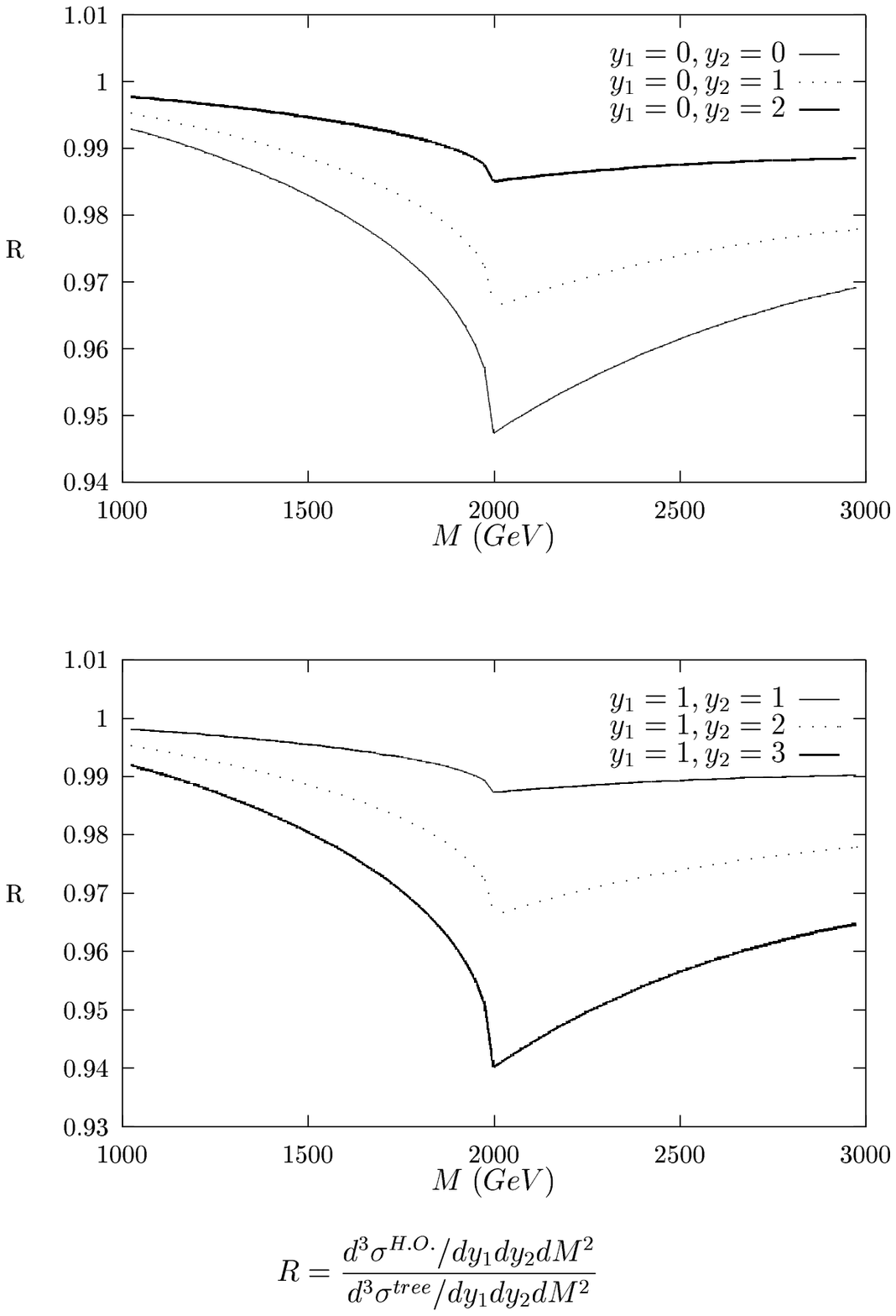}
\hss}
\caption[]{
Triple-differential cross section for $p \, p$ scattering
at $\sqrt{s}=14$ TeV, as a function of the two-jet invariant mass $M$, for
various choices of jet rapidities, calculated assuming $m_s=m_g=1$ 
TeV. Again, $R$ denotes the ratio
of the one-sparticle-loop-corrected cross section to the tree-level
cross section.}
\label{lhc1}
\end{figure}

We now use the above results to calculate one-loop corrections to
jet cross sections at large $E_T$ and two-jet invariant mass $M$, by 
convoluting the above subprocess results with parton distributions.
Since the higher-order sparticle corrections are only significant over a 
small range of parton subenergy  in the threshold region, it is convenient 
to start with
the triply-differential cross section as a function of $M$ and
the rapidities $y_1,y_2$ of the final-state jet pair. In terms
of the partonic squared matrix elements, this is given by
\begin{equation}
\frac{d^3\sigma}{dM^2dy_1dy_2}=\frac{1}{8\pi M^4}\sum_{ijkl}f_i(x_1)f_j(x_2)
 \left |{\cal M}_{ij}^{kl}(M^2,\hat{t})|^2+ |{\cal M}_{ij}^{kl}(M^2,\hat{u})|^2\right)
\label{triple} \end{equation}
where $f_i(x)$ is the parton distribution function for parton $i$,
${\cal M}_{ij}^{kl}$ is the matrix element for the scattering of parton $i$
and parton $j$ to partons $k,l$, and the jet rapidities $y_{1,2}$ are 
given by 
\begin{equation}
x_{1(2)}=\frac{M}{\sqrt{s}}e^{\pm(y_1+y_2)},
\label{rapidities}
\end{equation}
and
\begin{equation}
\hat{t}=-M^2-\hat{u}=-\frac{M^2}{1+\exp(y_1-y_2)}.
\label{hatthatu}
\end{equation}
To exemplify the results obtained with (\ref{triple}),
we consider a number of discrete choices of the
jet rapidities $y_{1,2}$, shown in Fig.~\ref{tev1} for
the Fermilab Tevatron, and in Fig.~\ref{lhc1} for the LHC.
We have used the parton-distribution functions from~\cite{MRS96},
and do not expect that using other parametrizations would
significantly affect our results.

\begin{figure}
\hbox to \hsize{\hss
\epsfxsize=0.69\hsize
\epsffile{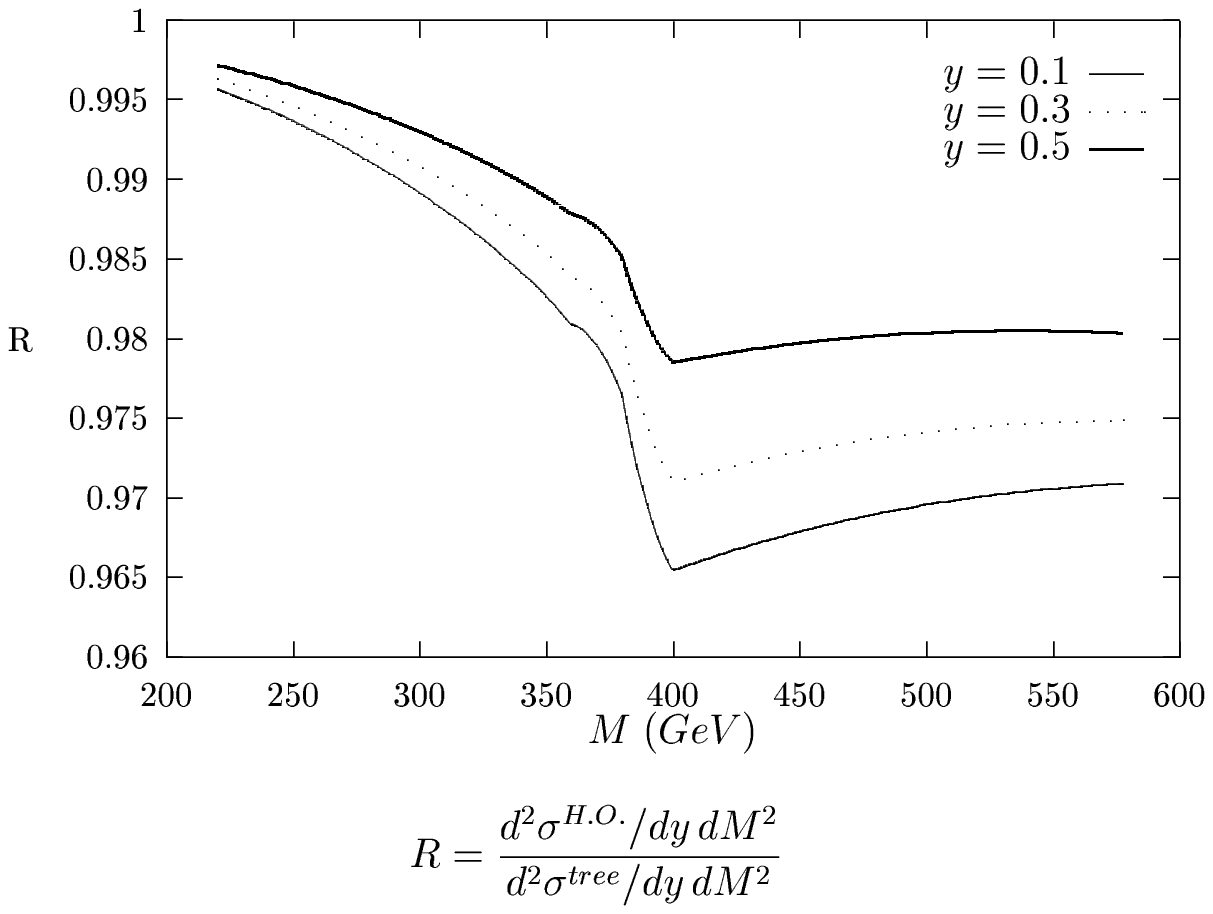}
\hss}
\caption[]{
Double-differential cross-section for $p \, \bar{p}$ scattering
at $\sqrt{s}=1.8$ TeV, as a function of the two-jet invariant mass $M$, for
various different jet rapidities, calculated assuming $m_s=m_g=200$ GeV. }
\label{tev3}
\end{figure}

\begin{figure}
\hbox to \hsize{\hss
\epsfxsize=0.69\hsize
\epsffile{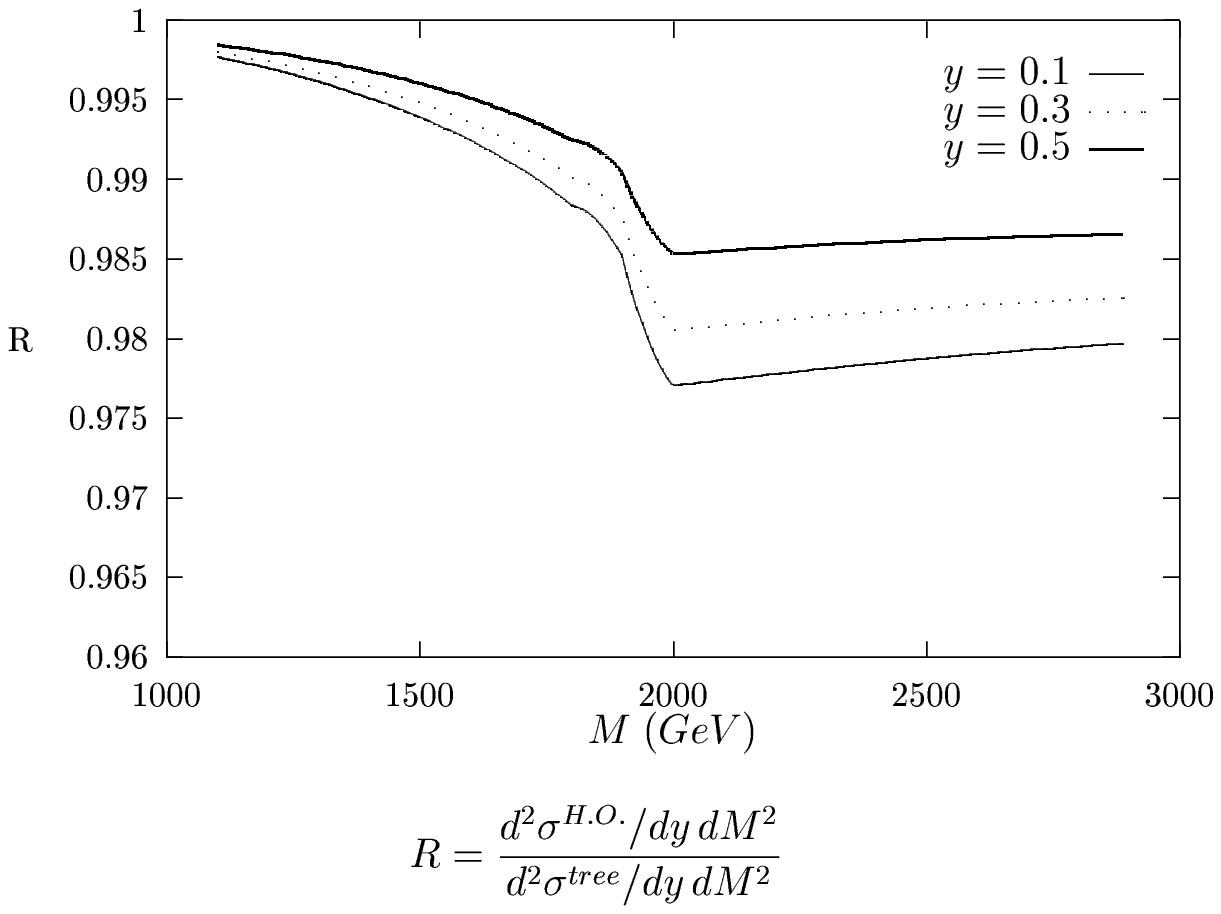}
\hss}
\caption[]{
Double-differential cross-section for $p \, p$ scattering
at $\sqrt{s}=14$ TeV, as a function of the two-jet invariant mass $M$, for
various different jet rapidities, calculated assuming $m_s=m_g=1$ TeV. }
\label{lhc3}
\end{figure}

The first striking feature is that the corrections are negative. This is 
because, over most of the ranges of parton momentum fractions $x_1, \, x_2$
studied, the subprocesses involving $t$-channel exchange dominate over the
annihilation processes. Indeed, recalling that we must symmetrize over the
partons in the final-state jets, we see from the tree-level
cross-sections given in~\cite{KEetal} that the 
subprocess cross section for
the scattering of two quarks with different flavours 
at any value of $s$ is {\it always}
an order of magnitude larger than the corresponding cross section for 
quark-antiquark annihilation, quite apart from the fact that
the parton distribution functions provide more
flux for this process. The second important 
feature of the plots shown in Figs.~\ref{tev1},\ref{lhc1}
is that, although characteristic cusps always appear at threshold, the 
reduction
of the differential cross-section due to the sparticle-loop corrections
is significant only for narrow ranges of rapidity pairs, $\{y_1,y_2\}$.
For this reason, these corrections are substantially washed out
if one integrates over one of the rapidities, in order to obtain the 
double-differential cross-section, in which the rapidity of only one of
the final-state jets is measured.
Nevertheless, for sufficiently small values of jet rapidity, there still
appears a cusp at threshold, as shown in  Fig.~\ref{tev3}
for the case of the Fermilab Tevatron 
collider. However, these cusps
are always somewhat diminished and broader than the 
previous cusps for the triple-differential cross sections.
This ``washing out'' effect is greater for the LHC shown in Fig.~\ref{lhc3},
where the dip in the threshold region is quite possibly too small
 to be observed.

\begin{figure}
\hbox to \hsize{\hss
\epsfxsize=0.69\hsize
\epsffile{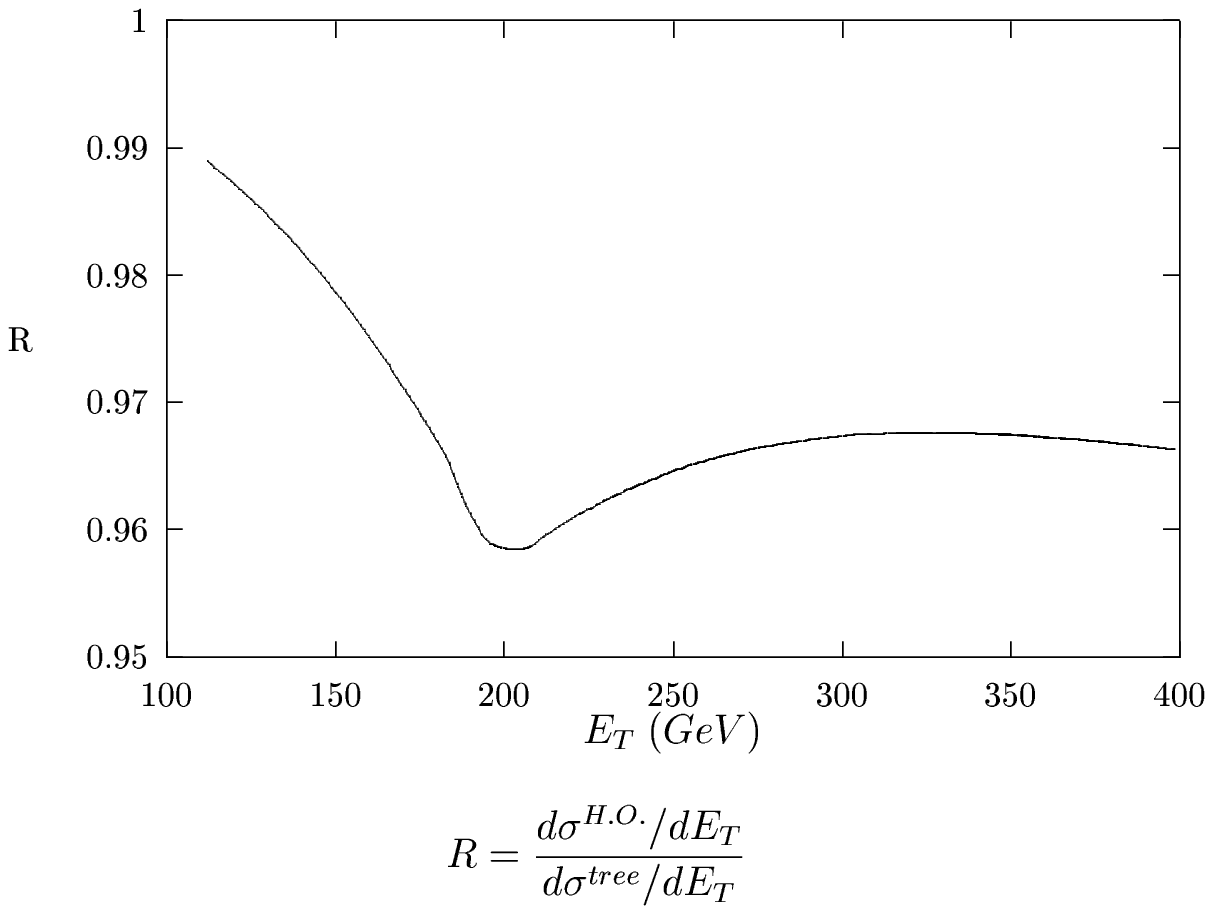}
\hss}
\caption[]{
Single-differential cross section for $p \, \bar{p}$ scattering
at $\sqrt{s}=1.8$ TeV, as a function of transverse energy  $E_T$, 
calculated assuming $m_s=m_g=200$ GeV. }
\label{tev4}
\end{figure}

\begin{figure}
\hbox to \hsize{\hss
\epsfxsize=0.69\hsize
\epsffile{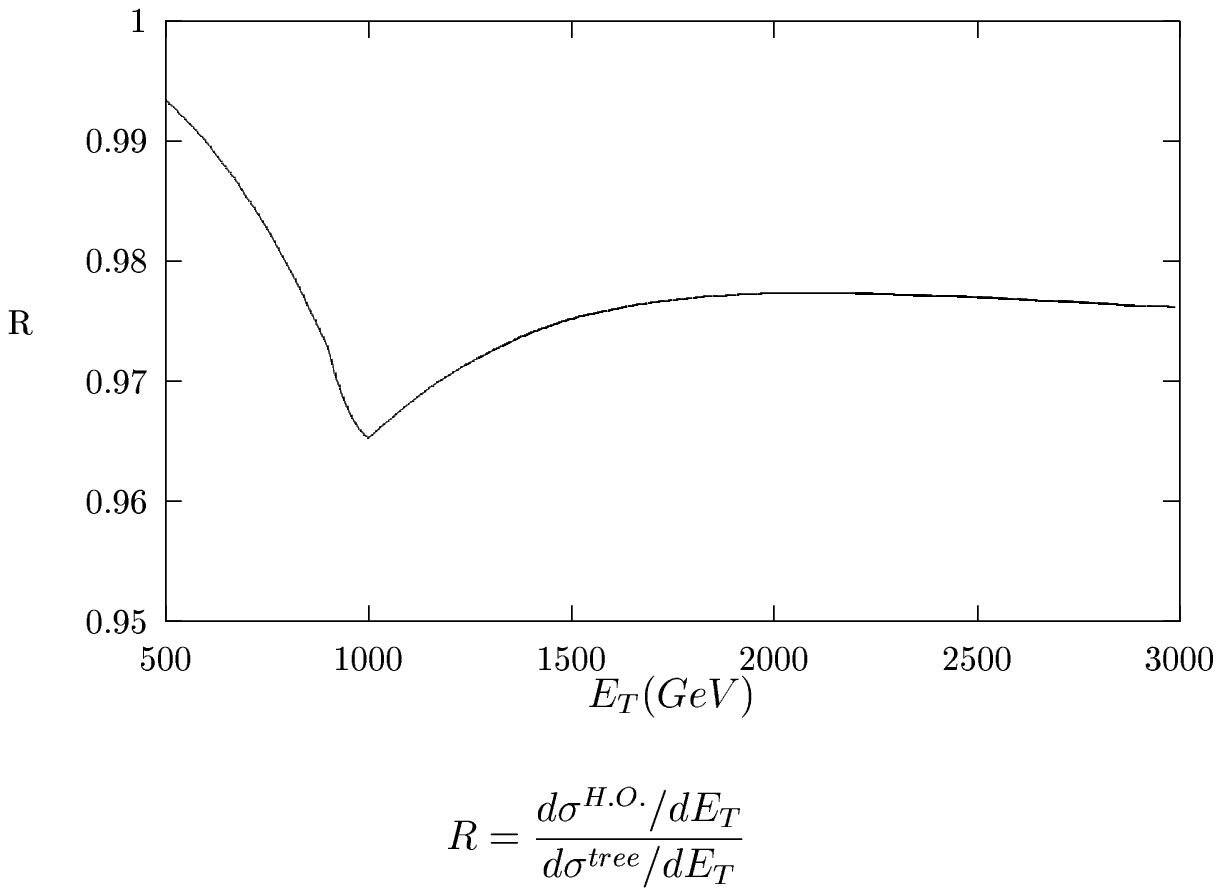}
\hss}
\caption[]{
Single-differential cross section for $p {p}$ scattering
at $\sqrt{s}=14$ TeV, as a function of transverse energy  $E_T$, 
calculated assuming $m_s=m_g=1$ TeV. }
\label{lhc4}
\end{figure}

This dilution is even more evident in the case of the single-differential 
cross section
as a function of the transverse energy $E_T$, as shown in Fig.~\ref{tev4}
for the case of the Tevatron. There is a broad dip in the cross section
for $E_T \simeq m$, but this is probably also too shallow to be
observable. This dip is again  shallower
 for the LHC large-$E_T$ cross section, shown in Fig~\ref{lhc4}.

In general, the sparticle-loop signal is clearer for sparticles of mass
$\simeq$ 200 GeV at the  Tevatron than for sparticles of mass $\simeq $ 1 TeV
at the  LHC. The reason for this is that smaller values of $m^2/s$ are
sensitive, on the average, to
smaller values of $x$, thereby sampling more of the 
gluon content of the incident hadrons. As we saw earlier, the sparticle-loop 
effects are smallest for purely gluonic scattering. 
It is, however, noteworthy that the net effect of
virtual-sparticle loops is to decrease the predicted cross section,
and can therefore not be used even as a partial explanation of any
unexpected rise in the large-$E_T$ differential cross section~\cite{CDF}.

\section{Conclusions}

We have presented in this paper complete one-sparticle-loop
corrections to the large-$E_T$ and large-$M$ cross sections
at high-energy hadron-hadron colliders, and used them to
present some numerical results for the Fermilab Tevatron collider 
and the LHC. We find that the
sparticle-loop effects are too small, and of the wrong sign, to
make any contribution to explaining the possible large-$E_T$
cross-section discrepancy reported recently~\cite{CDF}.
However, the fact that our calculated corrections exhibit cusps
at the sparticle threshold may encourage the hope that these
effects could be visible in the large-statistics data to be obtained
in the future at the Tevatron collider and the LHC.

As we have shown, these cusps are most noticeable in the triple-differential
jet cross section $d^3 \sigma / d M d y_1 d y_2$, and get progressively
more washed out as one integrates over one or both jet rapidities, or
if one plots the integrated large-$E_T$ cross section. It is for the
experimental collaborations to judge whether they will be able to obtain
the necessary statistics, and whether the systematic errors can be
controlled to the desired low level. In this paper we have not included
any allowance for experimental effects such as the initial transverse
momenta of the colliding partons, extra gluon radiation,
or the experimental resolution in the large-$E_T$ jet energies.

Evidently, we do not know where the squark and gluino thresholds
may be, nor whether they are coincident. The numerical results
presented in this paper have been for the optimistic case $m_s = m_g$,
and our threshold effects will be surely be spread out and further diluted 
to some extent if $m_s$ and $m_g$ are substantially different.
However, we would like to point out that looking for such a cusps
in differential cross sections is in principle a model-independent,
though indirect, way of looking for strongly-interacting sparticles.
The only vertices that enter our calculations are those
proportional to $\alpha_s$, and the squark and gluino decay vertices
do not enter. Thus, looking for the subtle effects we have
presented here is a strategy complementary to the direct
searches for sparticles decaying according to some particular
model scenario, which depends whether $R$ parity conserved or not, and on 
the spectrum of lighter sparticles. Also, the large cross section for
large-$E_T$ jets provides a window on large-mass physics that
may reach out to larger sparticle masses than direct searches in
specific decay modes with uncertain branching ratios.

Anybody interested in obtaining the code used to derive these results
should contact D.A.R., who will make it freely available. 

J.E. thanks the University of Melbourne School of Physics, and
D.A.R. thanks the CERN Theory Division, for hospitality during
the completion of this work.

\section*{Appendix}

\setcounter{equation}{0}
\renewcommand{\theequation}{A.\arabic{equation}}

In this appendix we list the VP~\cite{VP} functions used in
the text.

\subsection*{Tadpole function}
\begin{equation}
\int \frac{d^nk}{(2\pi)^n} \frac{1}{(k^2-m^2)}=\frac{i}{16\pi^2} A(m)
\label{one} \end{equation} 

\subsection*{Two-point Functions:}

\begin{equation}
\int \frac{d^nk}{(2\pi)^n} \frac{1}{(k^2-m_1^2)((k+p)^2-m_2)^2} 
=\frac{i}{16\pi^2} B_0(p^2,m_1^2,m_2^2).
\label{two1}
\end{equation}

\begin{equation}
\int \frac{d^nk}{(2\pi)^n} \frac{k^\mu}{(k^2-m_1^2)((k+p)^2-m_2)^2} 
=\frac{i}{16\pi^2} B_1(p^2,m_1^2,m_2^2) \, p^\mu.
\label{two2}
\end{equation}

\subsection*{Vertex functions:}
\begin{equation}
\int \frac{d^4k}{(2\pi)^4} \frac{1}{(k^2-m_1^2)((k+p_1)^2-m_2)^2
 ((k+p_1+p_2)^2-m_3^2)} 
=\frac{i}{16\pi^2} C_0
\label{three1}
\end{equation}

\begin{equation}
\int \frac{d^4k}{(2\pi)^4} \frac{k^\mu}{(k^2-m_1^2)((k+p_1)^2-m_2)^2
 ((k+p_1+p_2)^2-m_3^2)} 
=\frac{i}{16\pi^2} \left( C_{11} \, p_1^\mu+ C_{12} \, p_2^\mu \right),
\label{three2}
\end{equation}

\begin{eqnarray} 
\int \frac{d^nk}{(2\pi)^n} \frac{k^\mu k^\nu}{(k^2-m_1^2)((k+p_1)^2-m_2)^2
 ((k+p_1+p_2)^2-m_3^2)} &=&     \nonumber \\ & &
 \hspace*{-3in} \frac{i}{16\pi^2}
\left( C_{21} \, p_1^\mu p_1^\nu + C_{22}\,  p_2^\mu p_2^\nu
 + C_{23} (p_1^\mu p_2^\nu+p_2^\mu p_1^\nu ) + C_{24} g^{\mu\nu} \right), 
\label{three3}
\end{eqnarray}

\begin{eqnarray} 
\int \frac{d^nk}{(2\pi)^n} \frac{k^\mu k^\nu k^\rho }{(k^2-m_1^2)((k+p_1)^2-m_2)^2
 ((k+p_1+p_2)^2-m_3^2)} &=&     \nonumber \\ & &
\hspace*{-4in} \frac{i}{16\pi^2}
\left( \right. C_{31} \, p_1^\mu p_1^\nu p_1^\rho + C_{32}\,  p_2^\mu p_2^\nu p_2^\rho
 + C_{33} (p_1^\mu p_1^\nu p_2^\rho + p_1^\mu p_2^\nu p_1^\rho +p_2^\mu p_1^\nu p_1^\rho) 
 \nonumber \\ & &  \hspace*{-4in}
 + C_{34} (p_1^\mu p_2^\nu p_2^\rho + p_2^\mu p_1^\nu p_2^\rho
 +p_2^\mu p_2^\nu p_1^\rho)  \nonumber \\ & & -\hspace*{-4in}
+C_{35} (g_{\mu\nu}p_1^\rho + g_{\mu\rho}p_1^\nu + g_{\nu\rho}p_1^\mu)
+C_{36} (g_{\mu\nu}p_2^\rho + g_{\mu\rho}p_2^\nu + g_{\nu\rho}p_2^\mu)
\left. \right),
\label{three4}
\end{eqnarray}
where the functions $C_0, \, C_{ij}$ have in general the arguments
$(p_1^2,p_2^2,(p_1+p_2)^2,m_1^2,m_2^2,m_3^2)$. 
However, in our case we always have $p_1^2=p_2^2=0$, so we suppress 
the first two arguments.

\subsection*{Box functions:}
\begin{eqnarray}
\int \frac{d^4k}{(2\pi)^4}\frac{1}{(k^2-m_1^2)((k+p_1)^2-m_2^2)
((k+p_1+p_2)^2-m_3^2)((k-p_4)^2-m_4^2)} 
 &= & \nonumber \\ & & \hspace*{-3.2in}
\frac{i}{16\pi^2}D_0(2p_1\cdot p_2,2 p_2 \cdot p_4,m_1,m_2,m_3,m_4),
\label{four1}
\end{eqnarray}
\begin{eqnarray}
\int \frac{d^4k}{(2\pi)^4}\frac{ \ k^\mu}{(k^2-m_1^2)((k+p_1)^2-m_2^2)
((k+p_1+p_2)^2-m_3^2)((k-p_4)^2-m_4^2)}
 &= & \nonumber \\ & & \hspace*{-3.4in}
\frac{i}{16\pi^2}D_1(2p_1\cdot p_2,2 p_2 \cdot p_4,m_1,m_2,m_3,m_4,\mu),
\label{four2}
\end{eqnarray}
\begin{eqnarray}
\int \frac{d^4k}{(2\pi)^4} \frac{ \ k^\mu k^\nu}{(k^2-m_1^2)((k+p_1)^2-m_2^2)
((k+p_1+p_2)^2-m_3^2)((k-p_4)^2-m_4^2)}
 &= & \nonumber \\ & & \hspace*{-3.6in}
\frac{i}{16\pi^2}D_2(2p_1\cdot p_2,2 p_2 \cdot p_4,m_1,m_2,m_3,m_4,\mu,\nu),
\label{four3}
\end{eqnarray}
\begin{eqnarray} 
 \int \frac{d^4k}{(2\pi)^4}\frac{\ k^\mu k^\nu k^\rho}{(k^2-m_1^2)
((k+p_1)^2-m_2^2) ((k+p_1+p_2)^2-m_3^2)((k-p_4)^2-m_4^2)}
 &= & \nonumber \\ & & \hspace*{-3.8in}
\frac{i}{16\pi^2}D_3(2p_1\cdot p_2,2 p_2 \cdot p_4,m_1,m_2,m_3,m_4,\mu,\nu,\rho), 
\label{four4}
\end{eqnarray}
\begin{eqnarray}
\int \frac{d^nk}{(2\pi)^n}\frac{ \ k^\mu k^\nu k^\rho k^\sigma}{(k^2-m_1^2)
((k+p_1)^2-m_2^2)
((k+p_1+p_2)^2-m_3^2)((k-p_4)^2-m_4^2)}
 &= & \nonumber \\ & & \hspace*{-4.0in}
\frac{i}{16\pi^2}D_4(2p_1\cdot p_2,2 p_2 \cdot p_4,m_1,m_2,m_3,m_4,\mu,\nu,
\rho,\sigma).
\label{four5}
\end{eqnarray}
For the sake of compactness, we do not write these tensor expressions out in
terms of the vectors $p_1 \cdots p_4$, but refer the reader to~\cite{VP} 
for details.
\bigskip  

The exact forms of the functions $B_i,C_i,C_{ij}, D_i$ are given in~\cite{VP}.
The functions $A,B_0,B_1,C_{24},$ $C_{35},C_{36}, D_4$ are ultraviolet 
divergent, and therefore
should be calulated in $n=4-2\epsilon$ dimensions.
The pole parts of these functions are given by
the following expressions:
\begin{equation} {\rm P.P} \left\{ A(m) \right\} = \frac{m^2}{\epsilon}
\label{pol0} 
\end{equation}
 \begin{equation} {\rm P.P} \left\{ B_0(x,m_1,m_2) \right\} = \frac{1}{\epsilon}
\label{pol1} 
\end{equation}
\begin{equation} 
{\rm P.P} \left\{ C_{24}(x,m_1,m_2,m_3) \right\} = \frac{1}{4\epsilon} 
\label{pol2}
\end{equation}
\begin{equation} 
{\rm P.P} \left\{ C_{35}(x,m_1,m_2,m_3) \right\} = -\frac{1}{6\epsilon} 
\label{pol3}
\end{equation}
\begin{equation} 
{\rm P.P} \left\{ C_{36}(x,m_1,m_2,m_3) \right\} = -\frac{1}{12\epsilon} 
\label{pol4}
\end{equation}
 \begin{equation}{\rm P.P} \left\{ D_4(x,y,m_1,m_2,m_3,m_4,\mu,\nu,\rho,\sigma) \right\} =
 \frac{1}{24\epsilon} \left( g^{\mu\nu} g^{\rho\sigma}+
 g^{\mu\rho} g^{\nu\sigma} + g^{\mu\sigma} g^{\nu\rho} \right) 
\label{pol5} 
\end{equation}

\end{document}